\documentclass[utf8]{frontiersFPHY} 

\setcitestyle{square} 
\usepackage{url,hyperref,lineno,microtype,subcaption}
\usepackage[onehalfspacing]{setspace}

\def\keyFont{\fontsize{8}{11}\helveticabold }
\def\firstAuthorLast{M.E. G\'omez  {et~al.}} 
\def\Authors{M. E. G\'omez\,$^{1,*}$, S. Lola\,$^{3,4}$,  R. Ruiz de Austri,\,$^{4}$ and Q. Shafi\,$^{5}$ }
\def\Address{$^{1}$Departamento de Ciencias Integradas, Facultad de Ciencias Experimentales, Universidad de Huelva, 21071 Huelva, Spain\\
$^{2}$ Institute of Nuclear and Particle Physics, NCSR ‘Demokritos’, Athens, Greece\\
$^{3}$on leave from Department of Physics, University of Patras, Greece\\
$^{4}$ Instituto de F\'isica Corpuscular, IFIC-UV/CSIC, Valencia, Spain\\
$^{5}$Bartol Research Institute, Department of Physics and Astronomy, University of Delaware, Newark, DE 19716, USA }
\def\corrAuthor{1}

\def\corrEmail{mario.gomez@dfa.uhu.es}

\def\LSP {$\chi^0_1$}
\def\stau{$\tilde{\tau}_1$}

\def\stop{$\tilde{t}_1$ }

\begin{document}
\onecolumn
\firstpage{1}

\title[Confronting SUSY GUTs ...]{Confronting SUSY GUT with  Dark Matter, Sparticle Spectroscopy and Muon $(g-2)$} 

\author[\firstAuthorLast ]{\Authors} 
\address{} 
\correspondance{} 

\extraAuth{}

\maketitle

\begin{abstract}

\section{}
We explore the implications of LHC and cold dark matter
searches for supersymmetric particle mass spectra  in two different
grand unified models with left-right symmetry, $SO(10)$ and
$ SU(4)_c \times SU(2)_L \times SU(2)_R$ (4-2-2). We identify characteristic
differences between the two scenarios, which imply distinct correlations
between experimental measurements and the particular structure of the
GUT group. The gauge structure of 4-2-2 enhances significantly the
allowed parameter space as compared to $SO(10)$, giving rise to
a variety of coannihilation scenarios compatible with the LHC data, LSP
dark matter and the ongoing muon g-2 experiment.

\tiny
 \keyFont{ \section{Keywords:}Grand Unification, Sypersymmetry, Dark
   Matter, LHC, sparticle spectroscopy} 
\end{abstract}

\section{Introduction}

In recent years, LHC results \cite{higgs1, higgs2,CMSdat, ATLASdat}
and dark matter searches \cite{LUX, Akerib:2018lyp, Aprile:2015uzo,
  XENON1T, XENON1T_new, Aalbers:2016jon,PICO}
severely constrain some of the simplest standard model (SM)
extensions. Nevertheless, we know that we have to find a way
to go beyond the standard theory, which cannot accommodate massive neutrinos,
nor explain the observed baryon asymmetry of the universe
and recent cosmological observations
\cite{WMAP1,WMAP2,Ade:2013zuv,Ade:2015xua}.
These issues can be addressed by imposing further unification,
including grand unified theories and supersymmetry, which among others 
provides a natural candidate for dark matter \cite{DM-susy1,DM-susy2}. 

Here we consider two supersymmetric models with gauge
  unification at $M_{GUT}$, $SO(10)$ \cite{SO10G, SO10FM} and 
$ SU(4)_c \times SU(2)_L \times SU(2)_R$ (4-2-2) \cite{PS,PS-lr,PS2}.
We assume that at that scale the SUSY
  soft terms still preserve the group symmetry. 
This idea has been implemented in previous works for several 
GUTs \cite{EGLR,Okada:2013ija,
  coannih,Kowalska:2015zja,Kowalska:2014hza,Ellis:2016tjc}.  
Here, we focus on  the effects derived from the gauge structure 
of the two groups, while the sfermion mass terms remain universal,
with all matter being contained  in a single representation of the gauge group 
\footnote{This is the case in SO(10), but also in the LR symmetric
  4-2-2. A larger study relaxing the LR symmetrey in the 4-2-2 case is presented in 
  \cite{Gomez:2018zzw}}.
However, the soft higgs masses differ, since this sector is contained 
in different representations. The main difference between the two
groups arises in the gaugino sector: for $SO(10)$ it is natural to 
assume gauge mass universality, since this symmetry is broken to SM 
at the GUT scale. In the case of 4-2-2, however, the SM gauge couplings 
arise from combining broken and unbroken symmetries of the group,
allowing different GUT relations among the gaugino masses. 

 Even though the two models differ in just a single relation for the
  gaugino masses, this
results to vastly different  phenomenological predictions.
Among others, the possibility of small LSP masses in 4-2-2 enables
satisfying the relic density predictions due to coannihilations,
and gives rise to direct detection cross sections in the range of the current
experiments, as well as visible signals at the LHC.
It is also interesting to investigate whether in any of the two groups
the discrepancy between the experimental value of the anomalous
magnetic moment of the muon and the SM prediction  \cite{Davier:2010nc},
can be explained through supersymmetric contributions. 

Within this framework, we study the predictions for sparticle
spectroscopy in the two  scenarios, also requiring LSP dark matter
through coannihilations. We identify distinct
differences between the two groups, which result in direct correlations
between the experimental measurements and the group structure and
symmetries, which can be tested in future searches.

\section{GUT models and input parameter} 
\label{sec:3}
As in \cite{EGLR}, where $SO(10)$ has been compared with $SU(5)$ and 
flipped $SU(5)$, we assume that SUSY breaking occurs at a scale $M_X > M_{GUT}$, 
through a mechanism that 
generates flavour blind soft-terms.  Between the scales $M_X$ and
$M_{GUT}$, renormalisation and additional 
flavour symmetries may induce non-universalities
for soft terms that belong to different representations; on the
contrary, particles that
belong to the same representation have common soft masses.

The soft terms for the fields in an irreducible representation $r$ of the unification group are defined as multiples of a 
common scale $m_0$ as: 

\begin{equation}
m_{r}=x_r \, m_{0}, 
\end{equation}
while the trilinear terms are defined as: 
\begin{equation}
A_r = Y_r \,  A_0,  \;\;\; A_0=a_0 \, m_0 .
\end{equation}
Here, $Y_r$ is the Yukawa coupling associated to the $r$
representation.  We use the standard parametrisarion, with $a_0$ being
a dimensionless factor, which we consider as representation
independent. 
This is justified because the representation dependence is already
taken 
into account in the Yukawa couplings, and including a further factor
can be confusing.

Let us see how the above are applicable to the two gauge groups under
discussion:

$\bullet$ $SO(10)$

The simplest possibility arises within an $SO(10)$ GUT, in which 
all quarks and leptons are accommodated in the same
{\bf 16}  representation, leading to left-right symmetric
mass matrices.  We assume that the up and down 
higgses are in a pair of {\bf 10} representations.
This assignment determines sfermion mass
matrices and beta functions, and results in a common
mass for all sfermions
and two different higgs masses $m_{h_u}$  and  $m_{h_d}$
(thus identified with the NUHMSSM).

In addition to the CMSSM,  therefore, 
we introduce two new parameters $x_u$ and $x_d$ defined as: 
\begin{equation}
m_{16}=m_0, \;\;\; m_{H_u}=x_u \, m_{16}, \;\;\; m_{H_d}=x_d \, m_{16} . 
\end{equation}
Similarly, the $A$-terms are defined as: 
\begin{equation}
A_{16}=a_0 \cdot m_{0} .
\end{equation}

$\bullet$ 4-2-2 

The main features of the 4-2-2 that are relevant for our
discussion are summarised below. 
The 4-2-2 gauge symmetry can be obtained from a spontaneous breaking of
SO(10) by utilizing either the 54 dimensional or the 210 dimensional
representation,
allowing for some freedom of choice. The former case,
where $SO(10)$ breaks through a Higgs 54-plet
can be naturally combined with a left-right symmetry [34].
By contrast, this left-right symmetry is explicitly broken in
the latter case. 
Here, we will mostly focus on the left-right symmetric 4-2-2 model, 
which has the minimal number of free parameters and can be more directly
compared with $SO(10)$. In this case, the gaugino 
masses associated with $SU(2)_L$ and $SU(2)_R$ are the same,
while the gluino mass, associated with $SU(4)_c$, can 
differ. 
 
The main relations are therefore the following:

\begin{itemize}
\item Gaugino masses: The hypercharge
generator from 4-2-2 implies the relation
\begin{equation}
M_1=\frac{3}{5} M_2 + \frac{2} {5} M_3.
\end{equation}

\item Soft masses: All sfermions are accommodated in a 16
  representation, and have a common mass $m_{16} = m_0$. 
The Higgs fields are in a 10-dimensional representation 
with D-term contributions that result to 
$m^2_{H_{u,d}} = m^2_{10} \pm 2M_D^2$. In our notation, these values are:

\begin{equation}
r_u=\frac{m_{H_{u}}}{m_{16}}; \;\;\;\;\;\;\;\;\;\;
r_u=\frac{m_{H_{d}}}{m_{16}}; \;\;\;\;\;\;\;\;\;\;
\end{equation}
with $r_u<r_d$.

\end{itemize}

In our computations we assume a common unfication scale $M_{GUT}$ defined as the meeting point of the $g_1$ and $g_2$ gauge couplings. The GUT value for $g_3$ 
is obtained by requiring $\alpha_s(M_Z)=0.187$.  Above $M_{GUT}$ we assume a unification group that breaks at this scale.  SUSY is broken above $M_{GUT}$ 
by soft terms that are representation-dependent while preserving flavour blindness.
We perform a parameter space scan using as a guide  the representation
pattern at  the GUT scale for soft scalar terms. For this purpose, 
we extend the CMSSM universal scenario  through non-unified soft
terms, 
consistent with the representations of SO(10)
and 4-2-2. Even in their simplest versions, these scenarios result in soft term
correlations that enlarge the size of the parameter space that can is
compatible with  the neutralino relic density from Planck. 

We perform runs
with soft terms up to 10 TeV and a parameter range summarised below: 

\begin{eqnarray}
100 \text{GeV}\leq& m_0& \leq 10 \text{TeV} \nonumber\\
50 \text{GeV}\leq& M_{1}&\leq 4 \text{TeV} \nonumber\\
50 \text{GeV}\leq& M_{2}&\leq 4 \text{TeV} \nonumber\\
-10  \text{TeV}\leq& A_{0}&\leq 10 \text{TeV} \nonumber\\
2\leq& \tan\beta &\leq 65 \nonumber\\
-1.9\leq&  x_u&\leq 1.5 \nonumber\\
0\leq&  x_d&\leq 3.4 \nonumber\\
  x_u&\leq&x_d  
\end{eqnarray}
\label{sec:results}

\section{Experimental constraints and parameter space scan}
For our analysis  we use Superbayes
\cite{Bertone:2011nj,Strege:2012bt,Bertone:2015tza},
a package that
includes MSSM RGEs, relic density computations, phenomenological bounds and updated LHC 
bounds on SUSY particles, as well as phenomenological constraints
derived from b-physics. 
We further impose the constraint on the neutralino-nucleon 
cross section provided by the more recent Xenon-100 and LUX upper
limits. SuperBayeS-v2.0 performs a sample algorithm  using  
the code MultiNest v2.18 \cite{multinest}. 

It is well known that, if the required amount of relic dark matter is provided by neutralinos,
particular mass relations must be present in the 
supersymmetric spectrum. In addition to mass relations, we use the neutralino composition
to classify the relevant points of the supersymmetric parameter space.
The higgsino fraction of the lightest neutralino mass eigenstate is characterized by
the quantity
\begin{equation}
h_f \; \equiv \; |N_{13}|^2 + |N_{14}|^2 \, ,
\end{equation}
where the $N_{ij}$ are the elements of the unitary 
mixing matrix that correspond to the higgsino mass states.
Thus, we classify the points that pass the constraints discussed in Section 2 according 
to the following criteria:
\begin{itemize}
\item {\bf Higgsino \LSP  :}
\begin{flalign}
h_f >0.1, \;\;|m_A-2 m_\chi| > 0.1 \, m_\chi.
\label{criterio_higgsino}
\end{flalign}
In this case, the lightest neutralino is  higgsino-like and, as we discuss later, 
the lightest chargino $\chi^\pm_1$ is almost degenerate in mass with the \LSP.
The couplings to the SM gauge bosons are not suppressed and \LSP\ pairs have large 
cross sections for annihilation into $W^+ W^-$ and $ZZ$ pairs, which may reproduce the observed value
of the relic abundance. Clearly, coannihilation channels involving $\chi^\pm_1$ and 
$\chi^0_2$ also contribute. 

\item {\bf $A/H$ resonances:}
\begin{flalign}
|m_A-2 m_\chi|\leq 0.1 \, m_\chi.
\label{criterio_res}
\end{flalign}
The correct value
of the relic abundance is achieved thanks to $s$-channel annihilation,
enhanced by the resonant $A$ propagator. The thermal average $\langle \sigma_{ann}v\rangle$
spreads out the  peak in the cross section, so that neutralino masses for 
which $2m_{\chi} \simeq m_A$ does not exactly hold, can also undergo resonant annihilations.
\item {\bf $\tilde{\tau}$ coannihilations:}
\begin{flalign}
h_f <0.1,\;\;(m_{\tilde{\tau}_1}-m_\chi)\leq 0.1 \, m_\chi
\label{criterio_RR}
\end{flalign}
The neutralino is bino-like, annihilation into leptons through $t$-channel slepton exchange 
is suppressed, and coannihilations involving the nearly-degenerate \stau \ are necessary 
to enhance the thermal-averaged effective cross section.
\end{itemize}

 In the 4-2-2 model we get three additional types of coannihilation: 

\begin{itemize}
\item {\bf $\tilde{\chi}^+$ coannihilations:}

\begin{flalign}
h_f <0.1 ,\;\;(m_{\tilde{\chi}^+}-m_\chi)\leq 0.1 \, m_\chi.
\label{criterio_stop}
\end{flalign}

The lightest chargino  is light  and nearly degenerate with the bino-like neutralino.

\item{\bf $\tilde{g}$ coannihilations:}

\begin{flalign}
h_f <0.1,\;\;(m_{\tilde{g}}-m_\chi)\leq 0.1 \, m_\chi.
\label{criterio_stop}
\end{flalign}

The gluino is light  and nearly degenerate with the bino-like neutralino.

\item {\bf $\tilde{t_1}$ coannihilations:}
\begin{flalign}
h_f <0.1,\;\;(m_{\tilde{t}_1}-m_\chi)\leq 0.1 \, m_\chi.
\label{criterio_stop}
\end{flalign}
The \stop \ is light and nearly degenerate with the bino-like neutralino.
These coannihilations were found to also be present in the flipped
SU(5) model (but not $SO(10)$ or $SU(5)$).

\end{itemize}

We note that the LR symmetry in the scalar soft masses does not allow scenarios 
with $\tilde{\nu}-\tilde{\tau}-\chi$ coannihilations, which 
appear in the SU(5) scenarios of \cite{EGLR} or 
in the LR asymmetric 4-2-2.

%
%
%

\section{Planck compatible regions, muon anomalous magnetic dipolar moment  and Dark Matter Searches }

In $SO(10)$, the DM regions compatible with the Planck neutralino
relic density are similar to the CMSSM.  However, their parameter
space is enlarged due to the two independent higgs mass terms
(essentially reproducing the non-universal Higss CMSSM).  
Due to the additional freedom,
$\tilde{\tau}-\chi$ coanihilations (orange circles in the Figs.)  correspond to neutralino masses  below 650 GeV.  
Resonances of the pseudoscalar Higgs mediated neutralino annihilation
channels (brown crosses in the Figs.) are present in the entire range of $m_\chi$
under consideration, to be contrasted to the funnel-like area of the CMSSM.
Neutralinos with a large higgsino component are localized at values $m_\chi$ around 1 TeV. 

In 4-2-2, the GUT relation among the Higgsino masses results to wino-like
charginos and low mass gluinos that can coannihilate with the lightest
neutralino. Moreover, even if the sfermion masses are kept
universal at the GUT scale, it is possible to find models where light
stops can coannihilate with the neutralino.  In 4-2-2, therefore, the additional
freedom among the GUT values of the gaugino masses introduces 
three new kind of coannihilations that satisify the Planck neutralino
relic density requirements.

\begin{figure*}
\begin{center}
\vspace*{0cm}
\hspace*{-1.5cm}
\includegraphics*[scale=0.3]{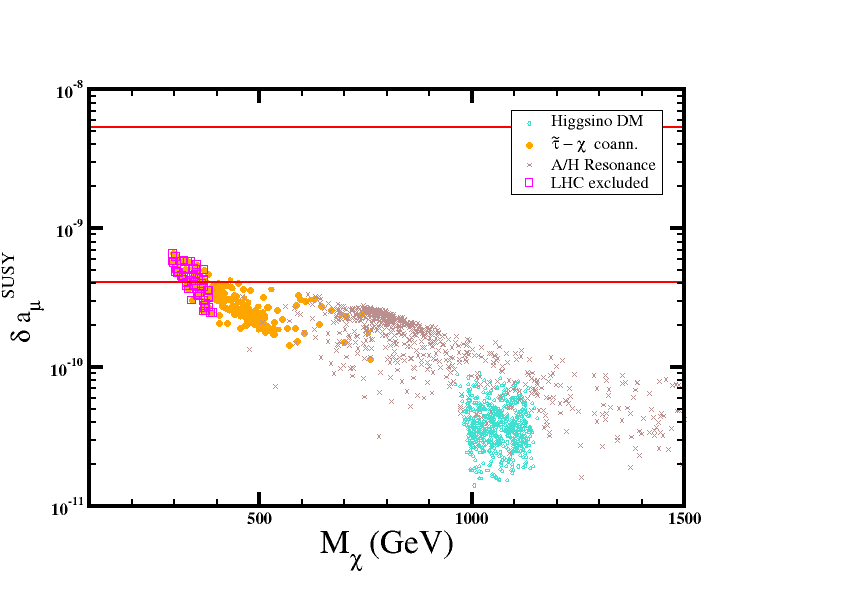}
\hspace*{-1cm}
\includegraphics*[scale=0.3]{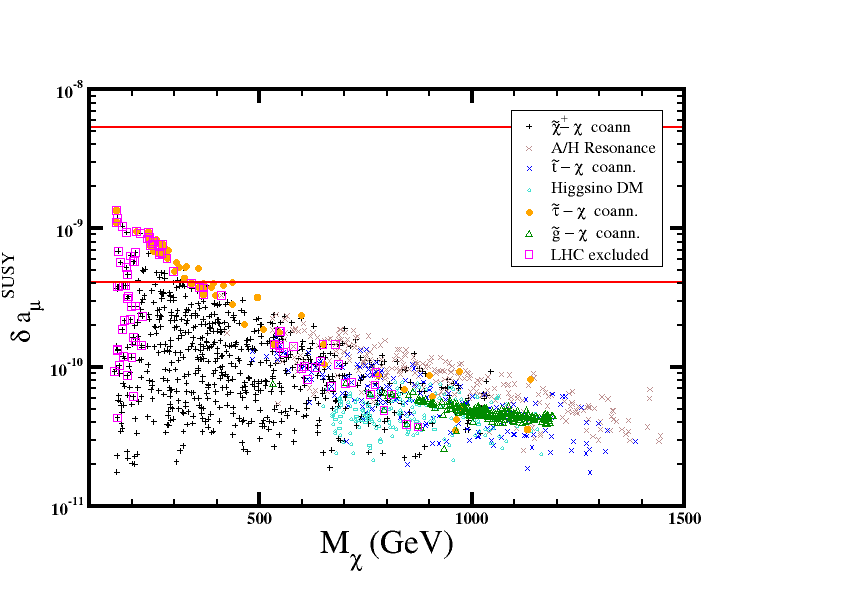}
\hspace*{-2cm}
\vspace*{-.5cm}
\caption{\it Prediction for  $\delta a_\mu^{SUSY}$ 
 versus $m_\chi$. The red
lines denote the 3-$\sigma$ bounds for the experimental discrepancy
of $a_\mu$ with respect to the SM prediction. The left plot
corresponds to SO(10) and the right to 
4-2-2. Different symbols and color codes are assigned to each class of
models  and this notation is maintained in the rest of the plots: 
Turquoise dots stand for Higgsino DM, black crosses for $\tilde{\chi}^{\pm}-\chi$ coannihilations, brown crosses for A/H resonances, blue crosses for $\tilde{t}-\chi$ coannihilations,
orange dots for $\tilde{\tau}-\chi$ coannihilations, and green triangles for $\tilde{g}-\chi$ coannihilations.}
\label{fig:delta_mlsp}
\end{center}

\end{figure*}

\subsection{Muon g-2}

The discrepancy between the experimental measurement of the muon $g-2$ and the respective SM prediction, $\delta a_\mu^{SUSY}=(28.7\pm8.2)\times 10^{-10}$, 
can be attributed to additional contributions from SUSY particles. However, for 
these contributions to be above the $3-\sigma$ level, sparticle masses
below $\sim 500$ GeV are required. 
By contrast, the experimental values for the Higgs mass
\cite{higgs1,higgs2}  can be accommodated in SUSY models with universal soft terms at the GUT scale with a heavy SUSY spectrum. Since in 4-2-2 unification the less restrictive gaugino mass
relations allow a lighter SUSY spectrum than in SO(10)
(also compatible with DM, especially for the models with chargino coannihilations), 
we expect that the study of $\delta a_\mu^{SUSY}$ favours this
unification group.  

In Figure \ref{fig:delta_mlsp} we display the prediction for $\delta a_\mu^{SUSY}$
versus the neutralino mass values. Although this will be discussed
later we include, already at this stage, the points
excluded by the LHC analysis.  We can see that the value of
 $\delta a_\mu^{SUSY}$ is always lower than the central value. In the
case of SO(10), we can see that just a few points can 
produce a significant contribution to the 
muon g-2 (however, these points are excluded by the LHC bounds). In
the case of 4-2-2 however, several points are compatible with a
contribution to the muon g-2; even though 
the respective region is small, it is compatible with chargino and
stau coannihilation regions 
that are not excluded by the LHC bounds.

\begin{figure*}[]
\begin{center}
\hspace*{-1.5cm}
\includegraphics*[scale=.3]{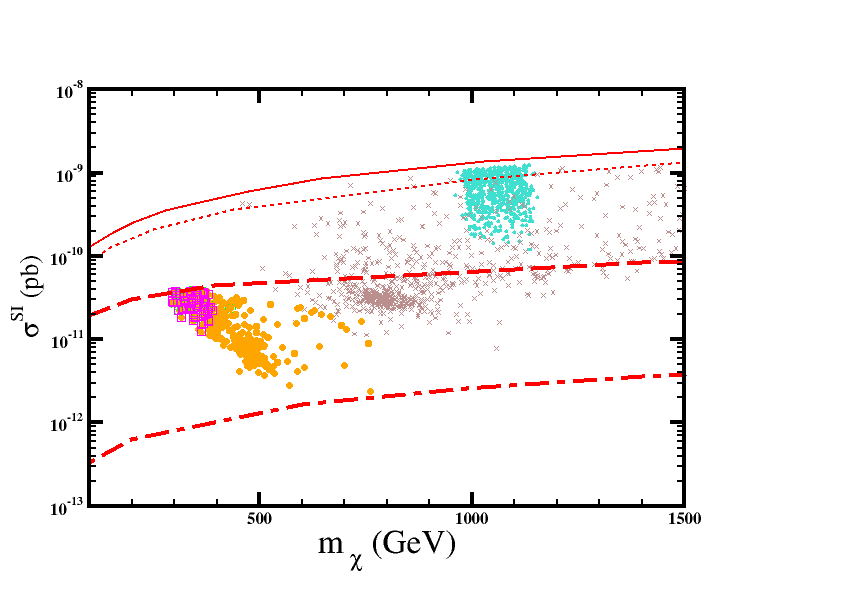}
\hspace*{-1 cm}
\includegraphics*[scale=.3]{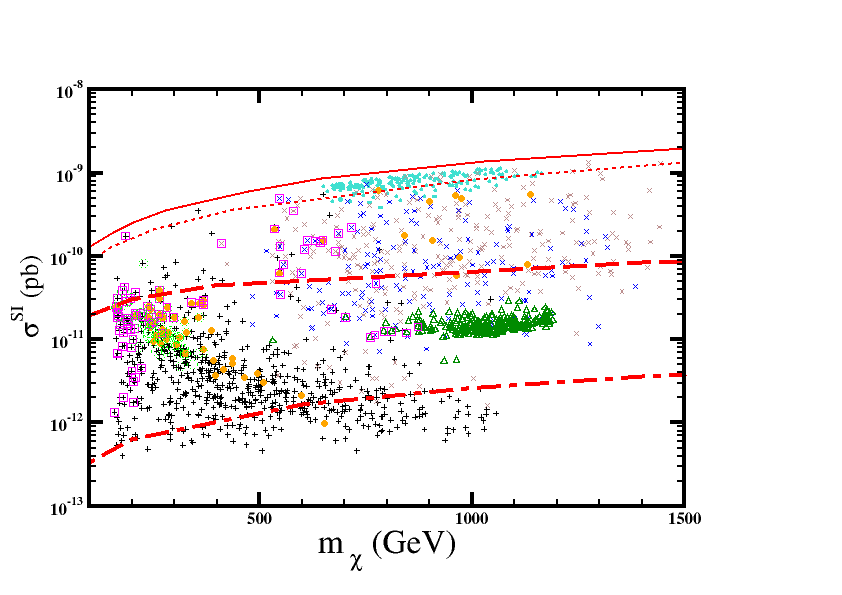}
\hspace*{-2cm}
\vspace*{-.5cm}
\caption{\it Scatter plot for the SI neutralino-nucleon cross
  section, for SO(10) (left) and 4-2-2 (right).
The red line corresponds to the  Xenon-1T bound \cite{XENON1T} while the dotted line includes their latest anouncement. The dash and the dot-dash lines correspond to the projected sensitivities from LZ\cite{Akerib:2018lyp} and DARWIN \cite{Aalbers:2016jon}. 
Symbols and color codes are similar to Fig.\ref{fig:delta_mlsp}}. 
\label{fig:simlsp}
\end{center}
\end{figure*}

\subsection{Dark Matter Searches}
The relation among different soft-terms will determine the composition
of the neutralino, which is important for its detection. 
In this respect, different GUTs result to different predictions,
which can tested in both direct and indirect detection experiments. 
Here we compare the spin independent (SI) neutralino-nucleon 
cross section of the models under
consideration, with the experimental bounds and prospects. 
Although spin dependent \cite{PICO} and indirect detection bounds
also exclude many SUSY models, 
the current and projected experiement for the SI neutralino-nucleon cross
section can test the predictions of the majority of the models considered here. 

Fig. \ref{fig:simlsp} is indicative of how a change on the gauge
unification conditions enables a significant enhancement of the
parameter space when passing from $SO(10)$ to 4-2-2. 
It is also possible to see how the direct detection bounds can impose 
important constraints on the parameter space. For instance, 
we can see that the latest update of the Xenon-1T bound \cite{XENON1T_new}
excludes many points with A-resonances and higgsino DM regions,
especially in the case of the 4-2-2 model.  But further than that,
Fig. \ref{fig:simlsp} shows that projected experiments can be
sensitive to most of the parameter space presented here.  In the
case of SO(10), DARWIN will cover the entire parameter space.  In the
case of 4-2-2,
certain areas of $\tilde{\chi}^\pm-\chi$ still remain below detection prospects.

\begin{figure*}[ht!]
\begin{center}
\hspace*{-1.5cm}
\includegraphics*[scale=.3]{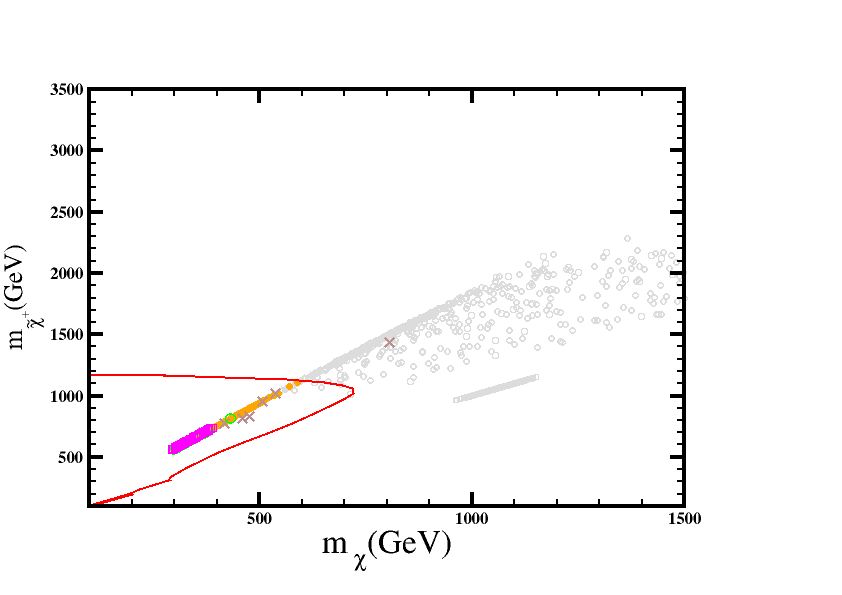}
\hspace*{-1cm}
\includegraphics*[scale=.3]{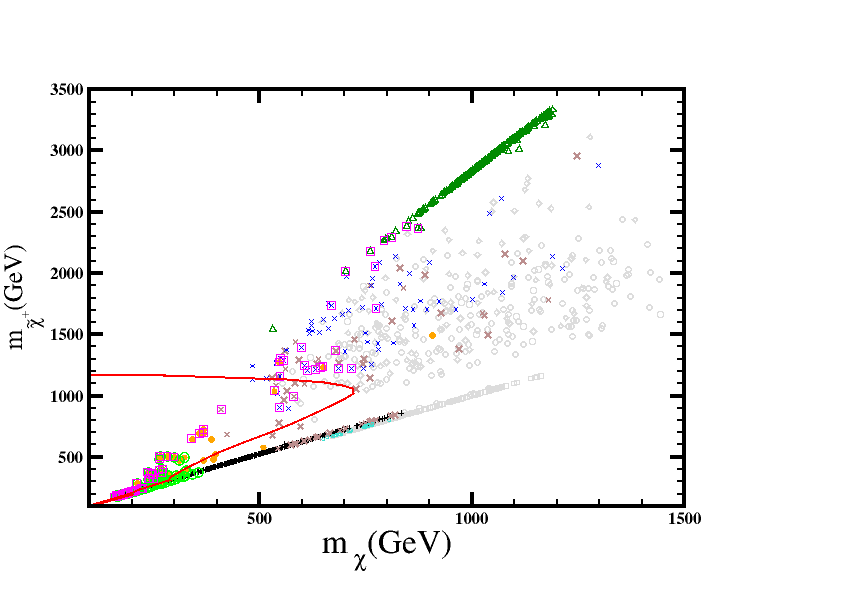}
\hspace*{-2cm}
\vspace*{-.5cm}
\caption{\it Chargino masses versus $m_\chi$ for the different 
  models of SO(10) (left) and 4-2-2 (right). The points 
that can be compared with the LHC bounds follow the same code as in
Figure ~\ref{fig:delta_mlsp}, while
the ones excluded are inside a purple square. Grey points correpond to
models that cannot be decomposed to SMS LHC signals (circles for
A-resonances, squares for higgsino DM and diamonds for stop
coannihilations). The red line correspond to the largest masses that
LHC can probe, according to the CMS and ATLAS public analysis.}
\label{fig:mcharmchi}
\end{center}
\end{figure*}

\section{LHC searches}

In this section we investigate the constraints imposed by the LHC on
the unified SUSY models under consideration. Each model can be
associated to a particular set of particle hierarchies and decays,
which are then compared with the generic data provided by the ATLAS
and CMS collaborations \cite{SMS_atlas, SMS_cms}. 
These comparisons are made with the help of Simplified Model Spectra
(SMS) which can be defined by a set of hypothetical SUSY particles 
and a sequence of products and decay modes that have to be
compared with those expected in our specific model. 
As a result, an individual check has to be done for every model, 
while, due to mismatches between the theoretical 
and the experimental results, it is not possible to
provide contour plots where one can easily see which mass ranges are excluded.
This task is simplified by using public packages like {\em
  Smodels-v1.1.1.}~\cite{Smodels},
which provides a powerful tool to perform a fast analysis of a large
number of models \cite{Kraml:2013mwa, Ambrogi:2017lov}. By using this
package, the theoretical models are decomposed in SMS and can be
contrasted with the existing LHC bounds if there is a match in the respective topologies.

\begin{figure*}[t!]
\begin{center}
\hspace*{-1.5cm}
\includegraphics*[scale=.3]{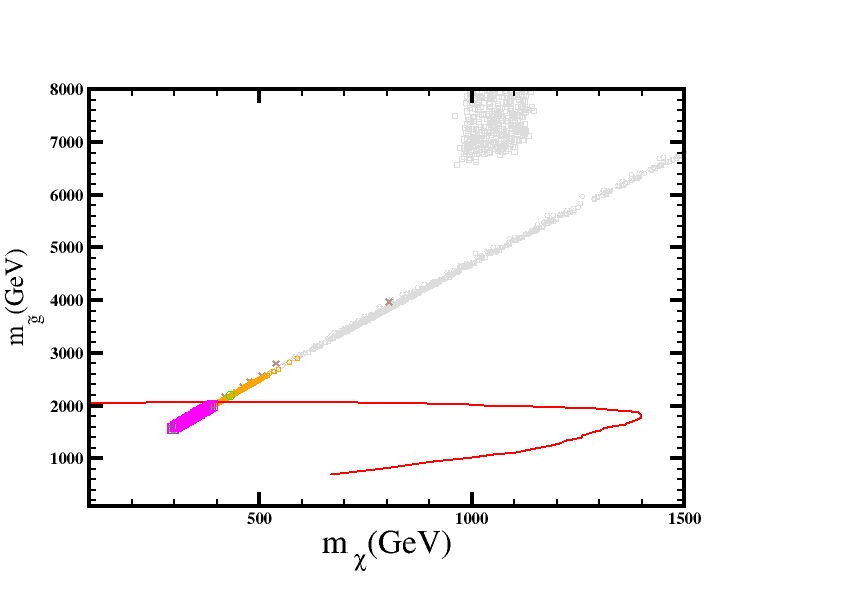}
\hspace*{-1.cm}
\includegraphics*[scale=.3]{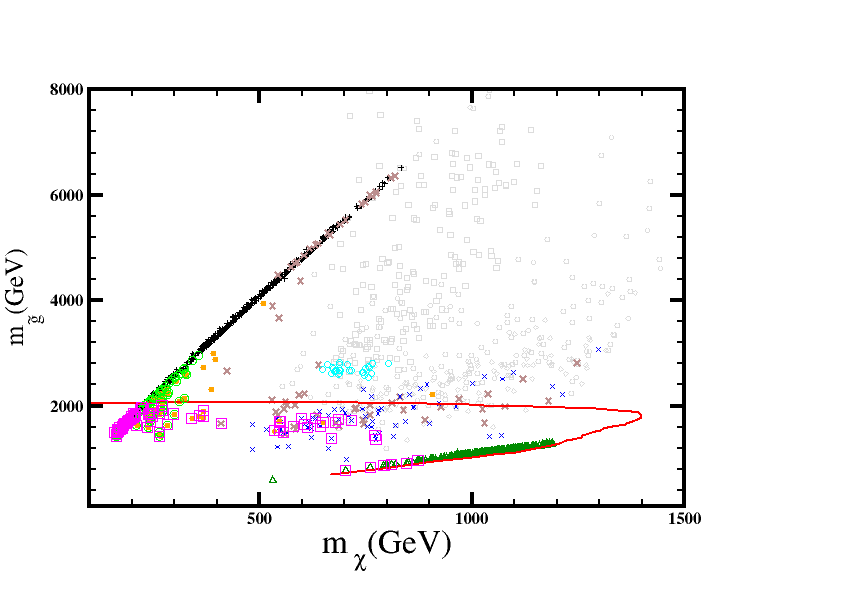}
\hspace*{-2cm}
\vspace*{-.5cm}
\caption{\it As in Fig.~\ref{fig:mcharmchi} for the $m_{\tilde{g}}$ versus $m_\chi$ scatter plot.}
\label{fig:mgmchi}
\end{center}
\end{figure*}

\begin{figure*}[]

\begin{center}
\hspace*{-1.5cm}
\includegraphics*[scale=.3]{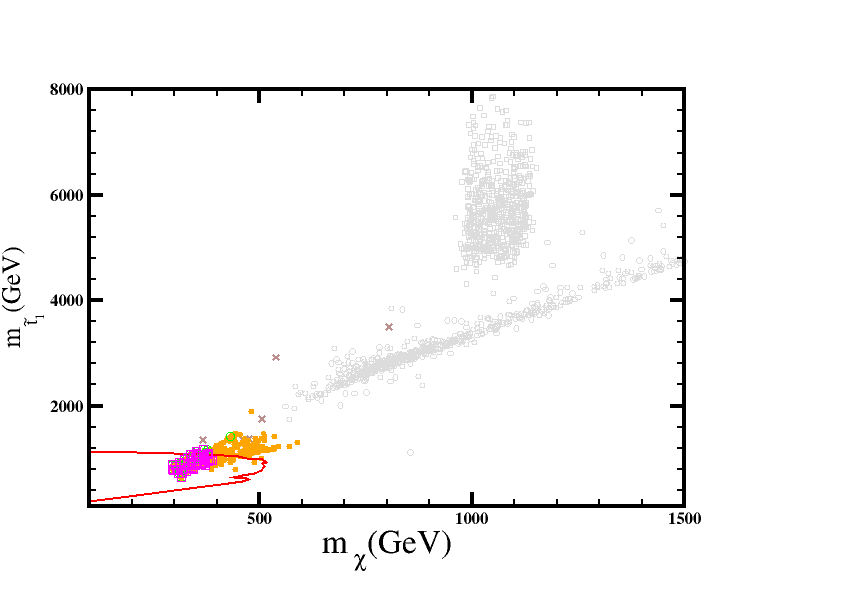}
\hspace*{-1cm}
\includegraphics*[scale=.3]{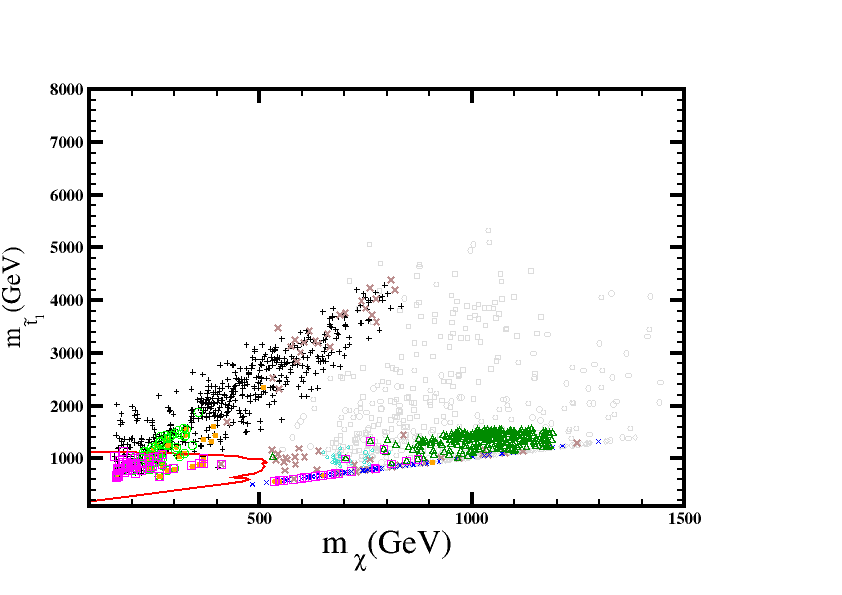}
\hspace*{-2cm}
\vspace*{-.5cm}
\caption{\it As in Fig.~\ref{fig:mcharmchi} for the $m_{\tilde{t}}$ versus $m_\chi$ scatter plot.}
\label{fig:mstopmchi}
\end{center}
\end{figure*}

\begin{figure*}[ht!]
\begin{center}
\hspace*{-1.5cm}
\includegraphics*[scale=.3]{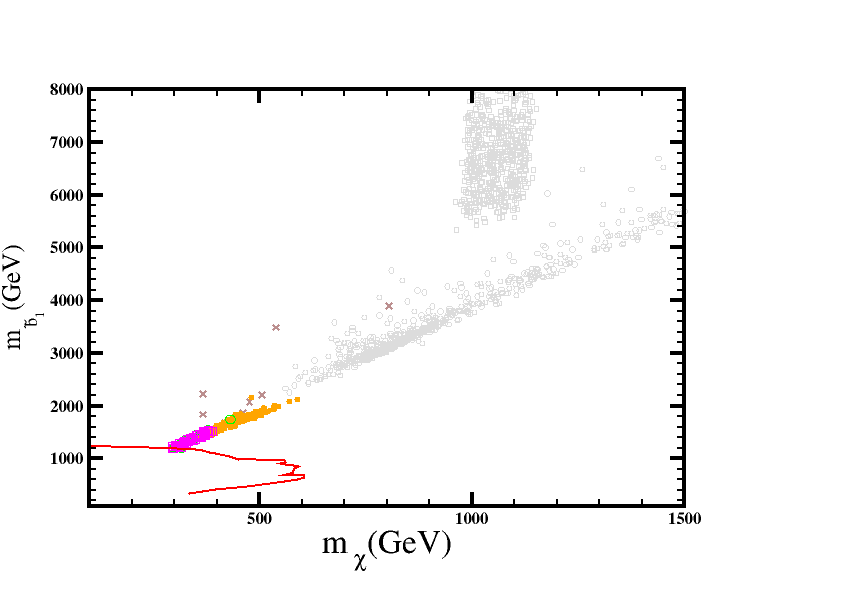}
\hspace*{-1 cm}
\includegraphics*[scale=.3]{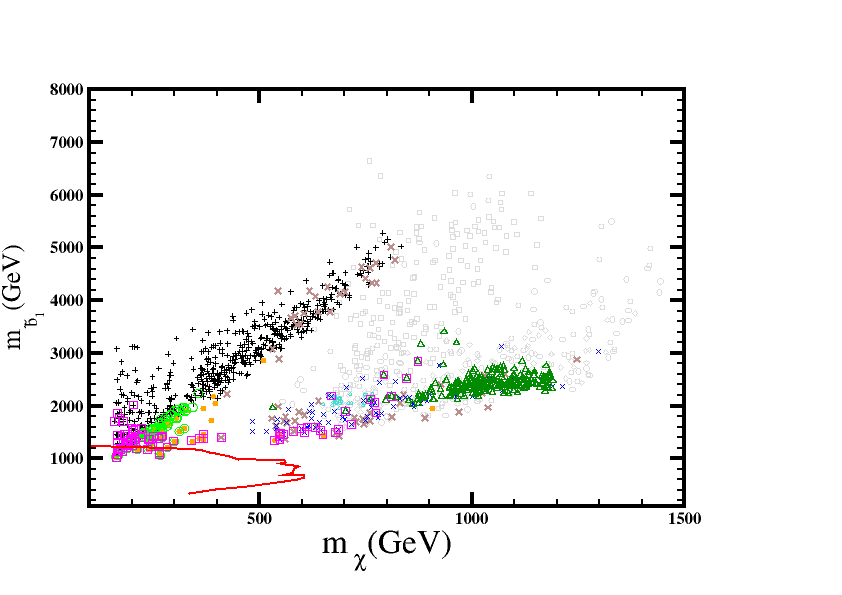}
\hspace*{-2cm}
\vspace*{-.5cm}
\caption{\it As in Fig.~\ref{fig:mcharmchi} for the $m_{\tilde{b}}$
  versus $m_\chi$ scatter plot.}
\label{fig:msbotmchi}
\end{center}
\end{figure*}

In Figs. \ref{fig:mcharmchi}~--\ref{fig:msbotmchi}
we classify the models as (i) the ones that can be compared with the
LHC data (either satisfying the bounds or being excluded) and (ii)
those that cannot be tested at the LHC; the latest are points that
either predict processes with very low coss sections or result in
topologies that are not tested at the LHC. 
For the points of category (i),
we follow the same notation as in previous
sections for points that satisfy the LHC bounds, 
and denote by magenta squares those excluded.
The points of category (ii) that cannot be tested, are drawn in grey.
For clarity, we display only points corresponding to A-resonant chanels (circles), 
higgsino DM (squares) and stop coannihilations (diamonds)
which lie at high mass areas in which the non-tested models dominate;
other classes of models lie in the same regions with the tested ones. 
The red solid boundary is obtained by combining
the simplified model bounds from LHC searches.
Since these models often do not apply to our particular cases
it cannot be considered as an exclusion line
(however, excluded points must be included in at least one of these contours). 
Nevertheless, it is useful to include this line for illustrative
purposes, since it gives 
an idea  of the range of masses explored at the LHC for every SUSY particle.

As we have already emphasized, despite the fact that the left-right
symmetric 4-2-2 unification differs from SO(10) by just one additional
gaugino mass, this gives rise to novel possibilities for DM models. 
Even similar classes of models that satisfy the  DM constraints, 
now correspond to a different range
of SUSY masses. This is due to the fact that the relic density
constraints are satisfied through coannihilations,
resonances or low values of the $\mu$ term, only 
in models with certain mass conditions 
(the areas of higgsino DM and A-resonnances have larger
values of SUSY partners in SO(10)). 
Consequently, the resulting LHC signals are very different in
the two groups. 

The SO(10) unified models are highly constrained by the LHC bounds. Most
of the points that Smodels can compare with the LHC bounds are in the 
stau coannihilation area, and a good part of them are already
excluded. These include a large majority of the models with a 
 muon g-2 contribution at the $3-\sigma$ level.  
No points on the Higgsino DM region can be tested at the LHC according
to the Smodels analysis and only a few points on the A-resoance area
can be reached, none of them excluded by the current bounds.  
The excluded points are affected mostly by the gluino  and stop
bounds. We can see in Fig.~\ref{fig:msbotmchi} that  
the predicted sbottom masses are outside the LHC accessible area.
The same happens with signals invloving quarks of the lighter generations. 

The 4-2-2 models have a richer structure with respect to experimental 
signatures at the LHC, than the ones arising from SO(10). In this case
we find points with higgsino DM and A-resonances in a wider spectrum
of masses; moreover, we find a larger number of points that Smodels
can decompose in signals that can be compared with the LHC bounds. In
Fig. \ref{fig:mcharmchi} we can see that models with chargino
coannihilations have a compressed spectrum that results to light
charginos compatible with the LHC bounds. However, we can see in
Fig. \ref{fig:mgmchi} that such points are excluded by the gluino
bounds. Despite that, some points predicting a relevant SUSY
contribution to the muon g-2 are not yet excluded  by the LHC. 
We can see in Fig. \ref{fig:mstopmchi} that bounds on stop searches
affect many of the excluded points, while only a few models predict
sbottom masses  that can be affected by LHC.  Processes involving 
first and second generation squarks are outside the scope of the current LHC bounds. 

\section{Conclusions}

In this work we have performed a comparative study of 
$SO(10)$ and
$ SU(4)_c \times SU(2)_L \times SU(2)_R$ grand unification
with respect to
the LHC sparticle mass spectra, cold dark matter and muon 
$g-2$ predictions.  Based on the remarkable complementarity 
between the LHC and dark matter searches, we 
show how the different patterns of  soft SUSY-breaking terms at
the GUT scale can be used to distinguish the two groups in
experimental searches. 

In particular, the gauge and symmetry breaking
structure of 4-2-2 enhances significantly the allowed parameter space
as compared to SO(10), and gives rise to three
additional coannihilation scenarios for dark matter
(chargino, gluino and stop coannihilations).
Even similar types of coannihilations that satisfy the  DM constraints, 
now correspond to a different range
of SUSY masses.
Moreover, areas where the discrepancy between the theoretical and
experimental values of  muon g-2 can be reduced via  a supersymmetric
contribution are identified. 

Gluino coannihilations are particularly important, since they are a
unique feature of 4-2-2
and do not appear in other GUT schemes. They are a direct
outcome  of the particular gaugino mass relations of the model that results in relatively light
gluinos. Chargino coannihilations are also found and, 
together with higgsino dark matter, are the most frequently encountered
scenarios. Stop coannihilations also arise (these can also appear in
flipped SU(5) \cite{EGLR}). In all cases, we get concrete predictions 
for the gaugino mass ratios that favour  the respected scenarios
and can be tested in future searches.

The overall message from the significant phenomenological differences between
two groups that share so many common features is clear: although no SUSY signal has
been found so far, there are still several alternative possibilities
to explore and the 4-2-2 group is one of them.

\section*{Acknowledgments}
The research of M.E.G. was supported by the Spanish MINECO, under grants FPA2014-53631- C-2-P and FPA2017-86380-P. R. RdA is supported by the Ram\'on y Cajal program of the Spanish MICINN, the Elusives European ITN project (H2020-MSCA-ITN-2015//674896- ELUSIVES), the “SOM Sabor y origen de la Materia” (PROMETEOII/2014/050) and Centro de excelencia Severo Ochoa Program under grant SEV-2014-0398. Q.S. acknowledges support by the DOE grant No. DE-SC0013880.
\bibliographystyle{frontiersinHLTH&FPHY} 
\bibliography{test}



\end{document}